\documentclass[aps,prl,amsmath,amssymb,superscriptaddress,showpacs,10pt,reprint]{revtex4-1}

\usepackage[dvips]{graphicx}
\usepackage[colorlinks=true]{hyperref}
\usepackage[normalem]{ulem}

\hypersetup{pdfpagemode = UseNone}
\usepackage{multirow}

\usepackage{braket}
\usepackage{dsfont}

\usepackage[rightcaption]{sidecap}  	
\sidecaptionvpos{figure}{t} 
\usepackage{enumitem}

\newcommand{\TC}{{T_\textsf{C}}}
\newcommand{\TQ}{{T_\textsf{Q}}}
\newcommand{\omegaC}{{\omega_\textrm{}}}
\newcommand{\omegaQ}{{\omega_\textrm{}}}

\newcommand{\betaC}{{\beta_\textsf{C}}}
\newcommand{\betaQ}{{\beta_\textsf{Q}}}

\newcommand{\betaX}{\beta_\textsf{X}}

\newcommand{\fX}{\textsf{X}}

\newcommand{\fQ}{\textsf{Q}}
\newcommand{\fD}{\textsf{D}}
\newcommand{\fC}{\textsf{C}}

\newcommand{\fQC}{\textsf{QC}}
\newcommand{\fQDC}{\textsf{QDC}}
\newcommand{\fR}{\textsf{R}}

\newcommand{\gibbs}{\hat{\mathcal{G}}}

\newcommand{\setup}[1]{{\textsf{#1}}}

\newcommand{\ie}{i.e.,~}

\newcommand{\Tr}{\mathrm{Tr}}

\newcommand{\nth}{{n_{\text{th}}}}

\newcommand{\remove}[1]{{\textcolor{red}{}}}

\usepackage{mathtools}

\newcommand{\dm}{\hat{\rho}}

\begin{document}
\title{Autonomous Maxwell's demon in a cavity QED system}
\author{Baldo-Luis Najera-Santos}
\affiliation{Laboratoire Kastler Brossel, Coll\`ege de France, CNRS, ENS-Universit\'e PSL, Sorbonne Universit\'{e}, 11 place Marcelin Berthelot, F-75231 Paris, France}
\author{Patrice A. Camati}
\affiliation{Universit\'e Grenoble Alpes, CNRS, Grenoble INP, Institut N\'eel, 38000 Grenoble, France}
\author{Valentin M\'etillon}
\author{Michel Brune}
\author{Jean-Michel Raimond}
\affiliation{Laboratoire Kastler Brossel, Coll\`ege de France, CNRS, ENS-Universit\'e PSL, Sorbonne Universit\'{e}, 11 place Marcelin Berthelot, F-75231 Paris, France}
\author{Alexia Auff\`eves}
\affiliation{Universit\'e Grenoble Alpes, CNRS, Grenoble INP, Institut N\'eel, 38000 Grenoble, France}
\author{Igor Dotsenko}
\email{igor.dotsenko@lkb.ens.fr}
\affiliation{Laboratoire Kastler Brossel, Coll\`ege de France, CNRS, ENS-Universit\'e PSL, Sorbonne Universit\'{e}, 11 place Marcelin Berthelot, F-75231 Paris, France}
\date{\today}

\begin{abstract}
We present an autonomous Maxwell's demon scheme. It is first analysed theoretically in term of information exchange in a closed system and then implemented experimentally with a single Rydberg atom and a high-quality microwave resonator. The atom simulates both a qubit interacting with the cavity, and a demon carrying information on the qubit state. While the cold qubit crosses the hot cavity, the demon prevents energy absorption from the cavity mode, apparently violating the second law of thermodynamics. Taking into account the change of the mutual information between the demon and the qubit-cavity system gives rise to a generalized expression of the second law that we establish and measure. Finally, considering the closed qubit-cavity-demon system, we establish and measure that the generalized second law can be recast into an entropy conservation law, as expected for a unitary evolution.
\end{abstract}
\maketitle

Back in the 19$^\textrm{th}$ century Rudolf Clausius has postulated that no spontaneous process exists whose sole result is the transfer of heat $\mathcal{Q}_h$ from a cold to a hot bath~\cite{Clausius1854}. Introducing the inverse temperatures, $\beta_c$ and $\beta_h$, of the cold and hot bath, respectively, this empirical result is captured by the formula 
\begin{equation}
	\Sigma \equiv \mathcal{Q}_h (\beta_h-\beta_c) \geq 0.\label{eq:Sthermo2}
\end{equation}
It was later realized that Eq.~\eqref{eq:Sthermo2} is one of the many expressions of the second law of thermodynamics (SLT),~$\Sigma$ being the entropy produced during the heat transfer. The SLT asserts that the total entropy of an isolated system can never decrease and stays constant for a reversible evolution, thus defining a thermodynamical arrow of~time. 

Later on, Maxwell pointed out a possible limitation of the SLT~\cite{Maxwellbook1975, Rex2017, Leff1990}. In contemporary terms, he envisioned a feedback mechanism, later dubbed Maxwell's demon, that would exploit the microscopic information on the molecules of a gas in order to establish a gradient of temperature, without any work expenditure. The paradox vanishes by acknowledging that information is a physical resource, that can be defined and quantified as correlations between the gas molecules and the demon's memory~\cite{Plenio2001,Plesch2014, Parrondo2015,Lutz2015}. While being consistent with the SLT, consuming these correlations allows performing tasks that would be impossible otherwise. Since these pioneering contributions, the Maxwell's demon has been experimentally exorcized in various setups~\cite{koski2014a, Koski2015, Camati2016, Vidrighin2016, Ciampini2017, Naghiloo2018, Masuyama2018, Wang2018}. These experiments made use of feedback loops, that require the processing of information at the classical level. Recently so-called autonomous Maxwell's demons got rid of this step, by exploiting information directly encoded on microscopic \cite{Koski2015} or quantum demons \cite{Cottet2017}.

Here, we report a theoretical description and an experimental realization of an autonomous Maxwell's demon scenario where information is exploited to transfer heat from a cold qubit to a hot harmonic oscillator(cavity). We first present a generalized SLT and derive  the conservation of entropy in this qubit-demon-cavity system. We then adapt our cavity QED setup to simulate and study this model. In drastic contrast with former experiments, the demon and the qubit are encoded on the same physical Rydberg atom, and the ensemble of subsystems (demon, cold body, and hot body) is closed. The whole protocol thus features a unitary, reversible evolution. While the entropy of the total system is constant, the entropy changes of the different subsystems are analyzed and measured as entropy production and consumption of correlations~\cite{Esposito2010, Ptaszy2019}.

We first focus on the general theoretical description of the underlying thermodynamic protocol. We study the transfer of heat $\mathcal{Q}_\fC$ from a qubit \setup{Q} to a cavity \setup{C}, that are initially prepared in thermal states $\gibbs_{\betaQ}$ and $\gibbs_{\betaC}$ with inverse temperatures $\beta_\fQ$ and $\beta_\fC$, respectively. This transfer is controlled by a demon~\setup{D}, the role of which is played by another qubit. We denote $\dm_\fX$, with $\fX \in\{\fQDC, \fQ, \fC, \fQC, \fD\}$, the density matrices of the total system and various subsystems, and  $\mathcal{S}_\fX \equiv \mathcal{S}[\dm_\fX] = -\Tr[\dm_\fX\ln\dm_\fX]$ their respective von Neumann entropies. The evolution of the joint qubit-demon-cavity system is unitary such that its entropy $\mathcal{S}_\fQDC$ remains constant in time. This entropy conservation law is consistent with the SLT as soon as unitary evolutions are considered reversible.

The demon's actions can be schematically split into a readout step and a feedback step. During the readout step, \setup{D} encodes information on the state of \setup{Q}. The amount of information owned by \setup{D} is quantified by the mutual information given by $\mathcal{I}_\textsf{QC:D} = \mathcal{S}_\fQC +  \mathcal{S}_\fD -  \mathcal{S}_\fQDC$. If the qubit is found in its ground state, the demon blocks the \setup{Q-C} interaction and, thus, the heat exchange between them: This is the feedback step, during which the demon state remains the same. Combined with the total entropy conservation, it yields $\Delta \mathcal{I}_\textsf{QC:D} =  \Delta \mathcal{S}_\fQC$. Here and in the following, $\Delta$ denotes the change of the corresponding quantity in the feedback step. In the absence of \setup{D}, there are no correlations and hence $\Delta \mathcal{S}_\fQC = 0$, as expected from a unitary evolution. Conversely, in the presence of \setup{D}, consuming correlations ($\Delta \mathcal{I}_\textsf{QC:D} <0$) allows one to lower the entropy of the \setup{QC} system at no work cost, which is the key mechanism at play in our experiment.

In a similar way, the \setup{QC} entropy change can be written as $\Delta \mathcal{S}_\fQC = \Delta \mathcal{S}_\fQ +  \Delta \mathcal{S}_\fC -  \Delta \mathcal{I} _\textsf{Q:C}$. Since the initial states of \setup{Q} and \setup{C} are thermal, their individual entropy changes are $ \Delta \mathcal{S}_\fX =  \mathcal Q_\fX \betaX - \mathcal{D} \left[ \dm_\fX  || \gibbs_{\betaX} \right]$, with $\fX \in \{\fQ,\fC\}$, $\dm_\fX$ the respective density matrices at the end of the feedback step and $\mathcal Q_\fX$ the heat absorbed by the system \setup{X}~\cite{SM}. They depend on the relative entropy $\mathcal{D}_\fX \equiv \mathcal{D}[\dm_\fX || \gibbs_{\betaX}] = -\Tr\left[\dm_\fX \ln \gibbs_{\betaX} \right] -  \mathcal{S}_\fX$, which is a strictly non-negative quantity measuring how much the system $\fX$ in state $\dm_\fX$ is far from a Gibbs state $\gibbs_{\betaX}$. From the two expressions for $\Delta \mathcal{S}_\fQC$ we get eventually
 \begin{equation}
    \mathcal{Q}_\fC (\betaC \!-\! \betaQ) \equiv \mathcal{Q}_\fC \delta\beta = \Delta \mathcal{I}_\textsf{QC:D} +  \mathcal{D}_\fQC.
    \label{eq:2nd_strong} 
\end{equation}
The term $\mathcal{Q}_\fC \delta\beta$ is reminiscent of the quantity appearing in Clausius inequality~\eqref{eq:Sthermo2}. From now on we shall dub it a {\it physical entropy production}. The relative entropy of the \setup{QC} system is defined as $\mathcal{D}_\fQC =  \mathcal{D} \left[ \dm_\fQ  || \gibbs_{\betaQ} \right] + \mathcal{D} \left[\dm_\fC  || \gibbs_{\betaC} \right] + \mathcal{I} _\textsf{Q:C}$, that can be recast as  $\mathcal{D}_\fQC =  \mathcal{D} \left[ \dm_\fQC  || \gibbs_{\betaQ} \otimes \gibbs_{\betaC} \right] \geq 0$ \cite{SM}. In the demon absence, Eq.~\eqref{eq:2nd_strong} simply reads $ \mathcal{Q}_\fC \delta\beta = \mathcal{D}_\fQC$ and the physical entropy production is positive as expected for a spontaneous process. Conversely, in the demon presence, the physical entropy production can reach negative values, revealing the role of information as a resource that allows for otherwise impossible dynamics. We refer to the inequality $\mathcal{Q}_\fC \delta\beta - \Delta\mathcal{I}_\textsf{QC:D}\geq 0$ as a generalized SLT~\cite{Sagawa2008}.

 \begin{figure}[t]
 		\includegraphics[width=\columnwidth]{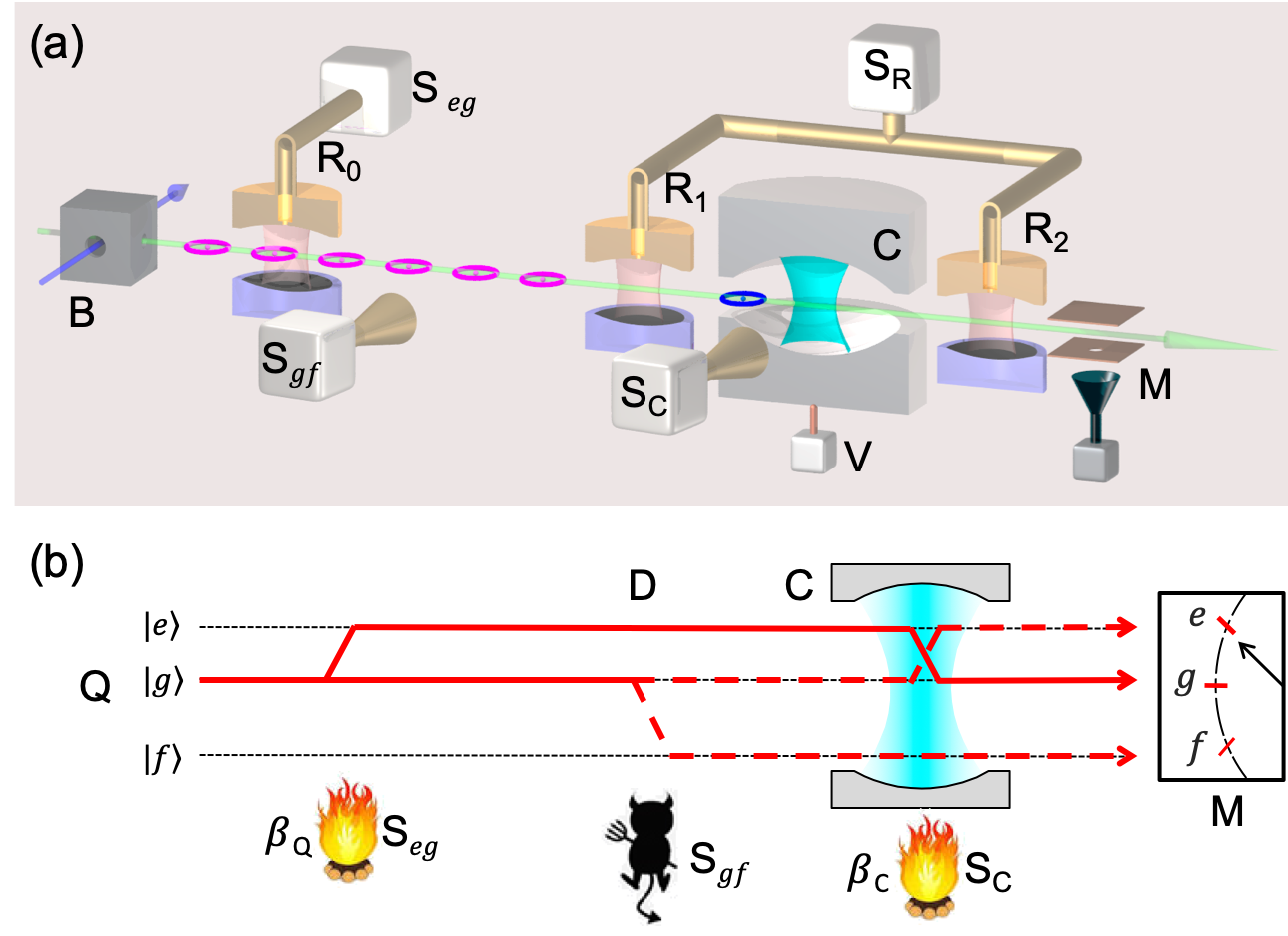}
 		\caption{Autonomous Maxwell's demon. (a) Scheme of the experimental setup with a microwave cavity (\setup{C}) and flying circular Rydberg atoms (blue toroid for the qubit atom, magenta toroids for probe probes). (b) Scheme of the experimental sequence. The cavity \setup{C}  is prepared in a thermal state at temperature $\betaC$ using microwave source \setup{S}$_\fC$. The three-level atom, initially in state $\ket g$, is prepared in a thermal state (temperature $\betaQ$) in the $\{\ket{g}\!, \ket e\}$ subspace using source \setup{S}$_{eg}$. The demon \setup{D} is realized by source \setup{S}$_{gf}$ resonant with $\ket f\!-\!\ket g$ transition and applied before \setup{Q} enters \setup{C}. The \setup{Q}-\setup{C} energy exchange is induced through adiabatic passage technique (\setup{Q}-\setup{C} detuning is controlled by electric field applied by voltage potential \setup{V}). The atomic state is measured by detector \setup{M}. The cavity state is reconstructed by the Ramsey interferometer (zones \setup{R}$_1$ and \setup{R}$_2$ fed by \setup{C}$_\fR$) with a sequence of probe atoms.}
 		\label{fig:setup}
	\end{figure}

Our experimental setup is depicted in Fig.~\ref{fig:setup}(a)~\cite{HarocheBook,Guerlin07}. It involves a high-finesse microwave cavity \setup{C}, made of two superconducting mirrors, and has a resonance frequency $\omega/2\pi= 51$~GHz and an energy relaxation time of $25$~ms at a $1.5$-K temperature. The source \setup{S}$_\fC$ injects a microwave field in \setup{C} by diffraction on the mirrors' edges. Individual circular Rydberg atoms with principal quantum number $50$ (level $\ket g$) are prepared in \setup{B} out of a velocity selected ($250$~m/s) beam of rubidium atoms. The mean atom number per atomic sample is about $0.04$, keeping the probability to have several atoms per sample negligible. The atomic transition frequency between $\ket g$ and the next higher circular level $\ket e$ is close to $\omega$. All circular states have a relaxation lifetime of about $30$~ms. The source \setup{S}$_{eg}$ drives the atoms between $\ket g$ and $\ket e$ in the zone \setup{R}$_0$. The atom-cavity interaction is controlled by means of the quadratic Stark shift in an electric field applied with potential \setup{V} across \setup{C}. The Rabi frequency of the resonant atom-cavity interaction is about $49$~kHz. The atomic levels are finally measured in the detector \setup{M} by state-selective field-ionisation. The detection efficiency is $0.5$ and the detection error, \ie the probability to erroneously attribute the atomic state, is $0.05$. The time of flight between \setup{B} and \setup{M} is $1.2$~ms, much shorter than all relaxation times. The Ramsey interferometer, composed of low-quality cavities \setup{R}$_1$ and \setup{R}$_2$ fed by the source \setup{S}$_\fR$, allows us to perform a quantum nondemolition measurement of the cavity photon number with a long sequence of probe atoms~\cite{Guerlin07,Metillon19}. The dispersive atom-cavity interaction for this measurement is tuned to realize a $\pi/2$ phase-shift per photon between atomic states $\ket g$ and $\ket e$.

We use this setup to verify the generalized SLT and the equality~\eqref{eq:2nd_strong}. The role of the harmonic oscillator is played by our microwave resonator \setup{C}. Conversely, both the qubit \setup{Q} and the demon \setup{D} are encoded on the same physical Rydberg atom. The simulation involves an auxiliary level of the Rydberg atom, namely a lower circular state $\ket f$  with principal quantum number $49$. The $\ket f\!-\!\ket g$ transition has a frequency of $54$~GHz and is driven by the source \setup{S}$_{gf}$. Eventually, the experiment leading to~\eqref{eq:2nd_strong} is recovered by using the following mapping between the three atomic levels and the logical states of \setup{Q} and \setup{D}:
\begin{equation}
	\ket e \!=\! \ket{1_\fQ}\otimes \ket{0_\fD}\!; \,\,      
	\ket g \!=\! \ket{0_\fQ}\otimes \ket{0_\fD}\!; \,\,      
	\ket f  \!=\! \ket{0_\fQ}\otimes \ket{1_\fD}\!. 
     \label{eq:logical}
\end{equation}
The fourth state of the \setup{QD} system, $\ket{1_\fQ}\otimes \ket{1_\fD}$, is never populated in the considered scheme and does not need to be taken into account~\cite{SM}. This mapping is bijective since it connects two sets of orthogonal states, introducing no information loss and not changing entropy. Therefore, all the quantities measured on the physical system can be unambiguously translated into thermodynamic quantities characterizing the simulated system.

Figure~\ref{fig:setup}(b) schematically presents the experimental sequence. The cavity \setup{C} is initially prepared in a thermal state $\gibbs_\betaC$ by means of \setup{S}$_\fC$. We perform $10$ microwave injections of $0.1$-ms duration each with the same amplitude, but random phase~\cite{SM}. The thermal field preparation is calibrated and verified by reconstructing the photon-number distribution $P(n)$ in $\gibbs_\betaC$ with a sequence of $800$ probe atoms. The estimated $P(n)$ is in excellent agreement with the Boltzmann thermal distribution providing us a thermal photon number $\nth$ and defining $\betaC=(\hbar\omegaC)^{-1}\ln[(1+\nth)/\nth]$. We set $\nth=0.63\pm0.04$~photons, corresponding to a temperature of $2.6\pm0.1$~K.

The qubit \setup{Q} is prepared in a thermal state $\gibbs_\betaQ$ by means of \setup{S}$_{eg}$. Since there is no phase in the state of \setup{C} and since the atomic state detection is phase-insensitive, we ignore the phase between $\ket g$ and $\ket e$. The qubit temperature in the $\{\ket{g}\!, \ket e\}$ subspace is defined by the population $p_e$ in the excited state as $\betaQ = (\hbar\omegaQ)^{-1}\ln [(1-p_e)/p_e]$ and is controlled by the duration of the \setup{S}$_{eg}$ pulse. We vary $\betaQ$ from large positive values (states close to $\ket g$) to large negative values (states close to $\ket e$) passing through $\betaQ=\betaC$ (mutual equilibrium state) and $\betaQ=0$ (infinite temperature with equal state population).

Then, the demon transfers the atomic population from $\ket g$ to $\ket f$. The duration of the \setup{S}$_{fg}$ pulse, applied before \setup{Q} enters \setup{C}, is pre-adjusted to reach the maximal population transfer of $0.95$ defining the demon read-out efficiency. Since level $\ket f$ does not couple to \setup{C}, the demon eliminates the possibility for the atom to extract energy from \setup{C}, while the probability to transfer the energy into \setup{C} when the atom is in $\ket e$ stays unchanged. When \setup{Q} enters \setup{C}, their energy exchange is made independent of the photon number of \setup{C} by using an adiabatic passage technique. The \setup{Q}-\setup{C} detuning is swept from $100$ to $-60$~kHz in $60\,\mu$s by means of the electric field applied across \setup{C}. The method ensures that an atom in $\ket e$ always injects a photon into \setup{C}, while an atom in $\ket g$ always absorbs a photon from \setup{C} if there was one (population transfer efficiency is better than $0.99$).  As expected, the overall demon's action has thus two effects: read-out (mapping of the state of \setup{Q} into the state of \setup{D}) and feedback (conditional protection of \setup{Q} from the following energy exchange with \setup{C}).  The demon's action is autonomous since it does not involve any information processing at the classical level.

The experimental sequence is repeated $25\,000$ times and the final state of \setup{C} is reconstructed separately for different detected states of \setup{Q} and \setup{D}~\cite{Metillon19}. In this way we obtain the probability $P(s_\fQ,s_\fD,n)$ for each joint \setup{QDC} state $\ket {s_\fQ} \otimes \ket {s_\fD} \otimes \ket n $, with $s\in\{0,1\}$ and $n$ photons in \setup{C}. This probability represents the diagonal elements (populations) of the state $\dm_\fQDC$ in the $\{\ket {s_\fQ} \otimes \ket {s_\fD} \otimes \ket n  \}$ basis. The state  $\dm_\fQDC$ is reconstructed for different qubit temperatures $\betaQ$~\cite{Note}. All measurements and state reconstructions are repeated with a reference experimental protocol which has no demon read-out.

 \begin{figure}[t]
 		\includegraphics[width=0.95\columnwidth]{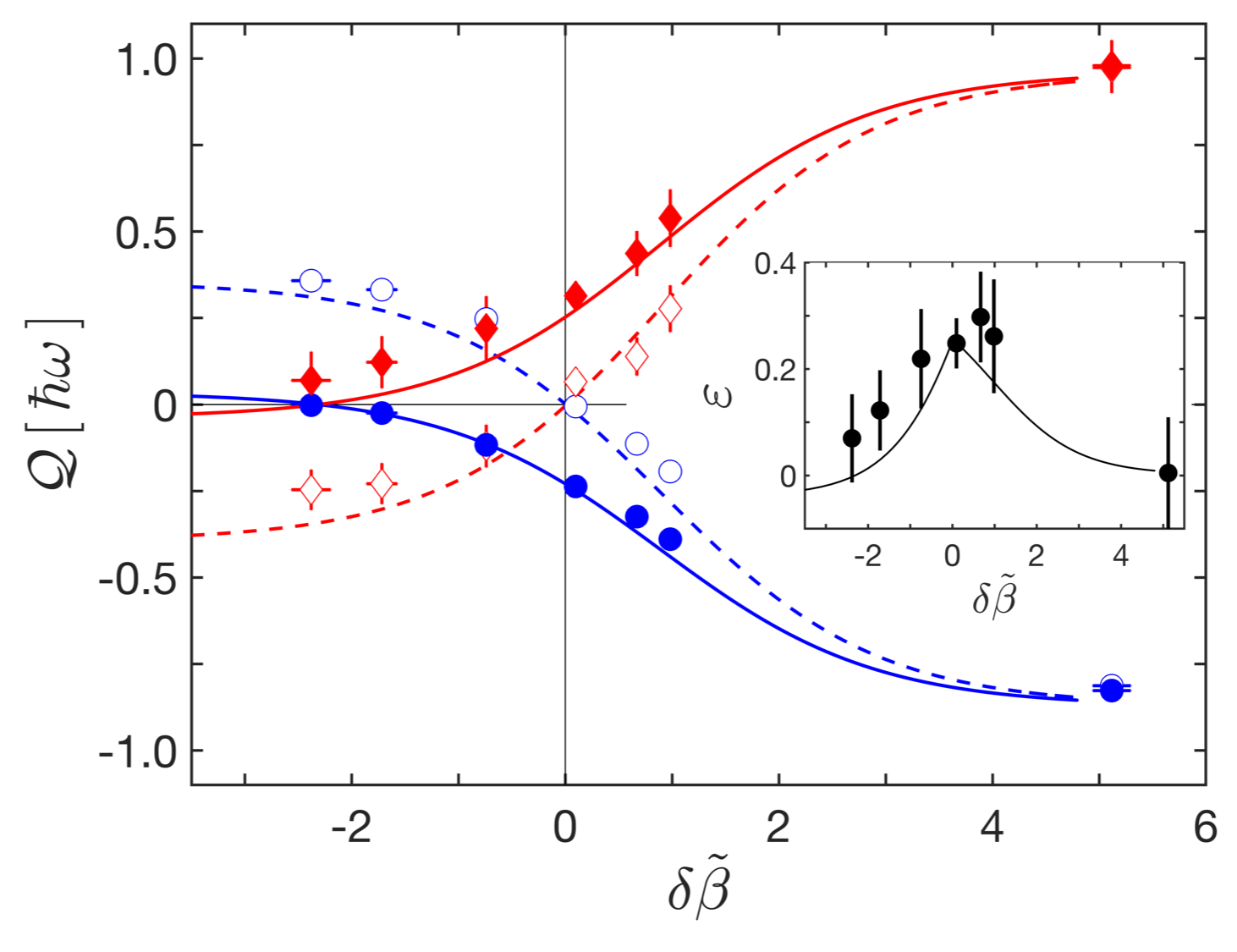}
 		\caption{Heat exchange versus relative inverse temperature $\delta\tilde{\beta}=1\!-\!T_\fC/T_\fQ$. Diamonds (circles) are the measured $\mathcal{Q}_\fC$  ($\mathcal{Q}_\fQ$). Lines are theoretical. Filled symbols with solid lines (open symbols with dotted lines) are in the presence (absence) of the demon. Vertical error bars are computed from the quantum state tomography. Horizontal error bars result from the finite accuracy of the set atomic temperature, especially for thermal states close to $\ket g$ and $\ket e$. Inset: the relative heat gain $\varepsilon$ of the cavity due to the demon, see text for details.}
 		\label{fig:heat}
	\end{figure}

\begin{figure*}[t!]
 		\includegraphics[width=\textwidth]{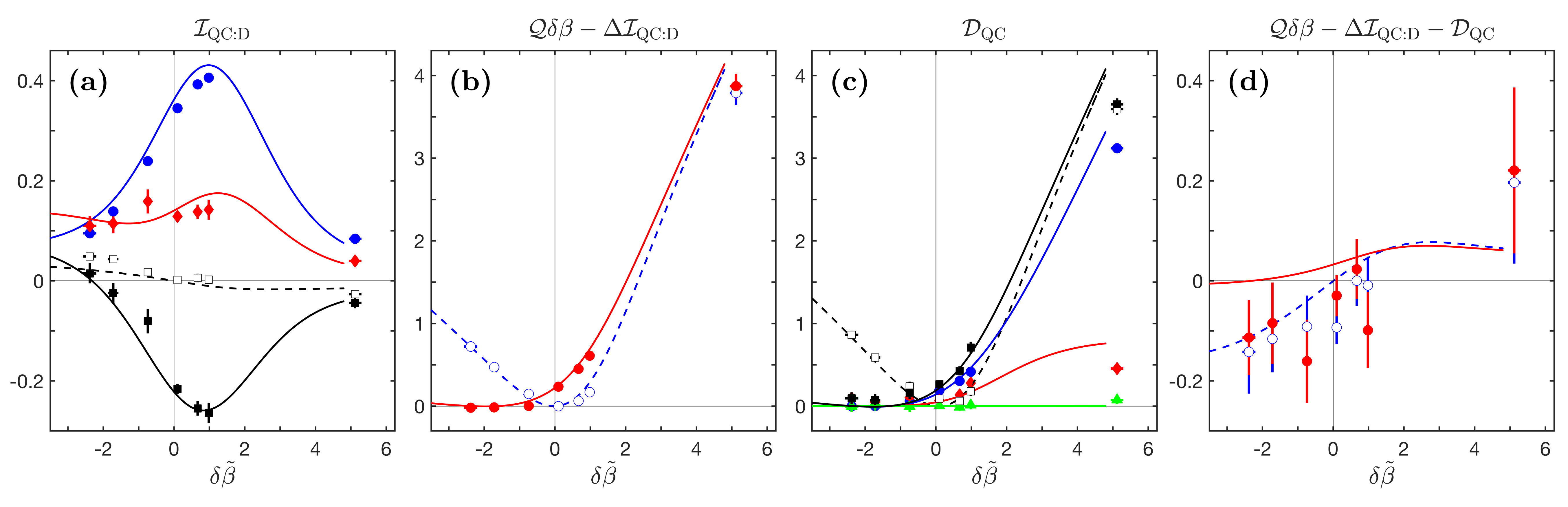}
 		\caption{Temperature dependence of thermodynamic quantities calculated in natural units of information (nats).
Symbols are measured results and lines are the corresponding theoretical evolutions. Filled symbols with solid lines (open symbols with dashed lines) are in the presence (absence) of \setup{D}.
(a)~Mutual information.  Blue, red, and black colours correspond to $\mathcal{I}_\textsf{QC:D}$ after read-out, to $\mathcal{I}_\textsf{QC:D}^\mathrm{fb}$ after feedback, and to their difference $\Delta \mathcal{I}_\textsf{QC:D}$, respectively. 
(b)~Generalized SLT. The entropy production $\mathcal{Q} \delta \beta$ is computed here using the qubit heat $\mathcal{Q}_\fQ$ which has a smaller experimental uncertainty than $\mathcal{Q}_\fC$.
(c)~Relative entropy. Blue (red) colours correspond to $\mathcal{D}_\fQ$ ($\mathcal{D}_\fC$), green - to $\Delta\mathcal{I}_\textsf{Q:C}$, black - to the sum $\mathcal{D}_\fQC$ of these three contributions. 
(d)~Total entropy change of the \setup{QDC} system. Points and line are the difference of the corresponding data in plots (b) and~(c). All error bars are calculated from the reconstructed states using an original method, developed in Refs.~\cite{ Six16,Metillon19,Perami19}, see also~\cite{SM}.}
 		\label{fig:results}
\end{figure*}

The \setup{Q-C} heat exchange $\mathcal{Q}$ is shown in Fig.~\ref{fig:heat} versus the relative qubit effective temperature $\delta\tilde{\beta} = 1-T_\fC/T_\fQ$. The heat received by \setup{C} is defined by $\mathcal{Q}_\fC=\hbar\omega\,\Delta \bar n$, where the change of the mean photon number, $\Delta \bar n$, is calculated from the photon-number distributions reconstructed with and without \setup{D}. Similarly, the heat received by \setup{Q} is defined by the change of the atomic population in $\ket e$ as $\mathcal{Q}_\fQ=\hbar\omega\,\Delta p_e$. The lines in Fig.~\ref{fig:heat} are theoretically computed taking into account the main experimental imperfections: the relaxation of \setup{Q} and \setup{C}, the limited demon efficiency, and the nonideal detection resolution~\cite{SM}. Figure~\ref{fig:heat} clearly shows that in the demon absence (open symbols and dashed lines) the hotter system always gives off heat to the colder one and that the physical entropy production is positive, $\mathcal{Q}_\fC\delta\beta>0$, as expressed by Eq.~\eqref{eq:Sthermo2}. At $\delta\tilde{\beta}=0$, \ie when \setup{Q} and \setup{C} have the same temperature, the net heat exchange is zero. In the demon presence (filled symbols and solid lines), however, the natural direction of heat flow can be reversed. Since \setup{D} is designed to block the energy transfer from the cavity to the qubit, \setup{Q} always gives off heat while \setup{C} always absorbs heat, irrespective of their temperatures. The sign change in $\mathcal{Q}$ for $\delta\tilde{\beta}<-2$ is mainly due to the limited demon efficiency providing the atom with a small probability to enter the cavity in level $\ket g$ and, thus, to absorb energy from~\setup{C}.

We characterize the performance of the demon by the heat gain, that we define as $\varepsilon = (\mathcal{Q}_\fC^\mathrm{d}-\mathcal{Q}_\fC^\varnothing) / \hbar\omega$. Here, $\mathcal{Q}_\fC^\mathrm{d}$ is the heat transferred from \setup{Q} to \setup{C} in the demon presence (shown by the red solid line in Fig.~\ref{fig:heat}). The quantity $\mathcal{Q}_\fC^\varnothing$ is the largest heat that can be transferred from \setup{Q} to \setup{C} in a protocol with no microscopic demon involved. The best classical strategy to maximize $\mathcal{Q}_\fC^\varnothing$, \ie knowing only the individual temperatures $\TQ$ and $\TC$, is to bring \setup{Q} and \setup{C} in contact if $\TQ>\TC$ ($\mathcal{Q}_\fC^\varnothing$ is then given by the red dashed line for positive $\delta\tilde{\beta}$ in Fig.~\ref{fig:heat}) and to inhibit their interaction if $\TQ\leq\TC$ ($\mathcal{Q}_\fC^\varnothing=0$ for $\delta\tilde{\beta}\leq0$). The inset in Fig.~\ref{fig:heat} presents the temperature dependence of the heat gain $\varepsilon$. For large positive $\delta\tilde{\beta}$, the transferred heat saturates close to its fundamental limit given by the energy of a single photon, $\hbar\omega$. Therefore, there is no more energy left for \setup{D} to further increase $\mathcal{Q}_\fC^\mathrm{d}$. On the other side, for large negative $\delta\tilde{\beta}$, \setup{Q} has no energy at all to transfer to \setup{C}, and thus neither method is able to extract the heat from  \setup{Q}. In the intermediate regime, however, the demon read-out (\ie the information on the microscopic state of the system) allows the extraction of more energy than in the classical case where only macroscopic information (\ie temperatures) is available.

Using $\dm_\fQDC$ reconstructed at each experimental step, \ie after initial state preparation, demon read-out, and feedback, we compute the quantities $\Delta\mathcal{I}_\textsf{QC:D}$ and $\mathcal{D}_\fQC$ entering Eq.~\eqref{eq:2nd_strong}. 
The results with and without the demon are shown in Fig.~\ref{fig:results}. There is no correlation between \setup{D} and \setup{QC} without read-out (open symbols in Fig.~\ref{fig:results}(a)). With the read-out activated, \setup{D} gets correlated with \setup{Q} (in blue). Part of this correlation is consumed during the feedback, but \setup{D} still keeps some non-consumed information on~\setup{C} after the feedback (in red). For an equal mixture of $\ket {0_\fQ}$ and $\ket {1_\fQ}$, \ie for $\delta\tilde{\beta}=1$, the mutual information between \setup{D} and \setup{Q} is the largest. Here, it is smaller than its absolute maximum value of $\ln(2)$ because of experimental imperfections mixing the atomic levels. In the limit $|\delta\tilde{\beta}|\gg 0$, the qubit's state tends to one of the pure states, $\ket{0_\fQ}$ or  $\ket{1_\fQ}$, having zero entropy. Thus, the mutual information $\mathcal{I}_\textsf{QC:D}$ being limited by the entropies of individual partitions also tends to zero. Note also, that for large negative $\delta\tilde{\beta}$, the atom is mostly in level $\ket g$, which can relax  to $\ket f$ or can be erroneously detected as $\ket f$. This effect creates small spurious  correlations between logical states of \setup{Q} and \setup{D}, resulting in non-zero values of $\mathcal{I}_\textsf{QC:D}^\mathrm{fb}$ for $\delta\tilde{\beta}\leq-2$, see the red curve in Fig.~\ref{fig:results}(a).

Figure~\ref{fig:results}(b) shows the generalized SLT. As expected, the consumption of the correlation $\mathcal{I}_\textsf{QC:D}$ can compensate the possible negative physical entropy production such that the quantity $\mathcal{Q} \delta \beta - \Delta\mathcal{I}_\textsf{QC:D}$ stays strictly non-negative. The relative entropy  $\mathcal{D}_\fQC$ is shown in black in Fig.~\ref{fig:results}(c). Its dominant term comes from the qubit, which starts in a thermal state and ends in an almost pure state $\ket{0_\fQ}$. This also implies that the \setup{QC} correlations are nearly zero and, thus, $\Delta\mathcal{I}_\textsf{Q:C}\approx0$. The relative entropy of \setup{C} is rather small and is maximal for $\beta_\fQ \!\rightarrow\! -\infty$ when the heat transfer is maximal (one photon). Figure~\ref{fig:results}(d) shows the entropy conservation of the total system as described by \eqref{eq:2nd_strong}. As expected, this equality is nearly valid both in the demon absence and presence. The residual deviation from zero comes from the non-negligible relaxations of \setup{Q} and \setup{C} and from the finite atomic state discrimination creating spurious correlations. 

In conclusion, we have analyzed our experiment like the transfer of heat from a cold body to a hot body by applying a mapping that identifies information as a resource to implement thermodynamically forbidden tasks. On the one hand, the mechanism at play is a global reversible unitary evolution coupling a qubit, a demon and a cavity while the total entropy is conserved. On the other hand, entropy changes of subparts of the joint system are identified with entropy production and information consumption, thus offering a striking experimental evidence that irreversibility and information are local physical concepts. By encoding information and working levels on the same physical system, our experiment sets a new landmark for autonomous Maxwell’s demons. Such, it is also interesting to study the thermodynamics of a qutrit-cavity system [ongoing work]. The resource here is the non-equilibrium distribution of the qutrit, that plays a similar role as information in the emerging framework of non-equilibrium Maxwell’s demons \cite{Sanchez19}.

\begin{acknowledgments}

We thank J.~P.~Paz for useful discussions. 
We acknowledge support by European Community (SIQS project) and by the Agence Nationale de la Recherche (QuDICE project).
P.~A.~C. acknowledges Templeton World Charity Foundation, Inc. This publication was made possible through the support of the grant TWCF0338 from Templeton World Charity Foundation, Inc. The opinions expressed in this publication are those of the author(s) and do not necessarily reflect the views of Templeton World Charity Foundation, Inc.
B.-L.~N.-S. and P.~A.~C. contributed equally to this work. 
\end{acknowledgments}


\begin{thebibliography}{10}
\bibitem{Note}
All states in the current work have no coherences and are thus diagonal in energy basis. However, for the sake of generality, we continue to use density matrices ($\dm$) instead of state populations ($\rho_{i,i}$) when describing state evolutions and calculating entropic quantities.
\end{thebibliography}
\end{document}